\documentclass[superscriptaddress,aps,tightenlines,amsmath,amssymb,floatfix,,raggedbottom,preprint,nofootinbib,nobalancelastpage]{revtex4-1}

\usepackage{graphicx}
\usepackage{subfig}
\usepackage{algpseudocode}
\usepackage{algorithm}

\begin{document}

\title{Hardware emulation of stochastic p-bits for invertible logic}
\author{Ahmed Zeeshan Pervaiz}
\email{apervaiz@purdue.edu	}      
\affiliation{School of Electrical and Computer Engineering, Purdue University, IN, 47907}

\author{Lakshmi Anirudh Ghantasala}
\affiliation{School of Electrical and Computer Engineering, Purdue University, IN, 47907}

\author{Kerem Yunus Camsari}
\affiliation{School of Electrical and Computer Engineering, Purdue University, IN, 47907}

\author{Supriyo Datta}
\email{datta@purdue.edu}      
\affiliation{School of Electrical and Computer Engineering, Purdue University, IN, 47907}
\date{\today}

\begin{abstract}
The common feature of nearly all logic and memory devices is that they make use of stable units to represent 0's and 1's. A completely different paradigm is based on three-terminal stochastic units which could be called ``p-bits’’, where the output is a random telegraphic signal continuously fluctuating between 0 and 1 with a tunable mean. p-bits can be interconnected to receive weighted contributions from others in a network, and these weighted contributions can be chosen to not only solve problems of optimization and inference but also to implement precise Boolean functions in an inverted mode. This inverted operation of Boolean gates is particularly striking: They provide inputs consistent to a given output along with unique outputs to a given set of inputs. The existing demonstrations of accurate invertible logic are intriguing, but will these striking properties observed in computer simulations carry over to hardware implementations? This paper uses individual micro controllers to emulate p-bits, and we present results for a 4-bit ripple carry adder with 48 p-bits and a 4-bit multiplier with 46 p-bits working in inverted mode as a factorizer. Our results constitute a first step towards implementing p-bits with nano devices, like stochastic Magnetic Tunnel Junctions.
\end{abstract}

\maketitle

\section*{Introduction}

Contemporary logic and memory devices are largely built from standard MOS (metal-oxide-semiconductor) transistors, but the possibility of alternative devices based on new materials and phenomena for both Boolean and non-Boolean computation has been discussed extensively (see for example ref.\cite{nikonov2015}). The common feature of nearly all such devices is that they make use of stable and deterministic units to represent 0's and 1's. A completely different paradigm is based on three-terminal stochastic units where the output is a random telegraphic signal $m_i(t)$ that continuously fluctuates between 0 and 1 and the mean value can be tuned with an analog signal $I_i(t)$ at the input terminal. In mathematical terms

\begin{equation}
\rm {m_i}(t) = {\rm{sgn}}\Big\{ \mathrm{rand(-1,1)} + \mathrm{tanh}\big({I_i}(t)\big)\Big\} 
\label{eq:psl3 p-bit}
\end{equation}

\noindent where rand($-$1,+1) represents a random number uniformly distributed between $-$1 and +1, while the retention time $\rm \tau_N$ of the p-bit is assumed large enough that memory of the last state $\rm m_i(t)$ has been lost. If the input is zero, the output $\rm m_i(t)$ takes on a value of $-$1 or +1 with equal probability. A negative input $I_i$ makes negative values more likely while a positive input makes positive values more likely.

\par Each such unit could be called a ``p-bit'' with an apparent similarity to ref.\cite{cheemalavagu2005}, and many such units can be correlated to perform useful functions by building an interconnected network where the analog input to the $\rm i^{th}$ p-bit consists of a bias $\rm h_i$ added to a weighted sum of the outputs $\rm m_j(t)$ of the other p-bits:

\begin{equation}
\rm {I_i}(t) = I_0\Big\{{h_i} + \sum_j {{J_{ij}}{m_j}(t)}\Big\} 
\label{eq:psl3 weight}
\end{equation}

\par We have recently shown that with a proper choice of the matrices $\rm \{h\}$ and $\rm [J]$, p-bit networks could be not only used to solve problems of optimization and inference \cite{behtash2016,sutton2017intrinsic} but also to implement precise Boolean functions in an invertible mode \cite{camsari2016,faria2017low}. 

\par This invertible operation of Boolean gates is a particularly striking characteristic very different from standard digital gates which provide a unique output in response to a set of inputs. This is also true of a Boolean gate implemented with p-bits, but it additionally provides all the inputs that are consistent with a given output. Even when there is no unique input, the gate fluctuates among the multiple allowed inputs.

\par The inverse operation is made possible by the bidirectional nature of the interconnection matrix $\rm [J]$ whereby both $\rm J_{ij}$ and $\rm J_{ji}$ are generally non-zero so that any two p-bits, say ``i'' and ``j'', influence each other, unlike standard digital logic with directed connections. A Boltzmann Machine (BM) \cite{ackley1985learning} with fully bidirectional connections, (all $\rm J_{ij} = J_{ji}$) , would put inputs and outputs on an equal footing. However, a BM would normally provide approximate answers without the kind of accuracy expected from digital logic. A directed network of bidirectional BM's, on the other hand, has been shown to provide a striking combination of digital accuracy and logical invertibility.

\par These demonstrations of accurate invertible logic are intriguing, but they are based on purely software implementations of Eqs.~(\ref{eq:psl3 p-bit},\ref{eq:psl3 weight})  and it is natural to ask whether real hardware implementations of these equations would preserve these striking properties. It is well-established that software implementations of unrestricted Boltzmann Machines need to be serially updated to ensure proper operation and convergence\cite{suzuki2013chaotic,hinton2007boltzmann}. In software, this is enabled by control flow statements such as``for-loops'' that make each update one by one, negatively impacting performance. How does this carry over to hardware implementations? In our hardware emulation, the serial updating of p-bits comes naturally without any peripheral control circuity. This is due to the asynchronous operation of p-bits that result from natural time delays between p-bits. In simulation p-bits are assumed identical, but how will inevitable process variations in real p-bit retention time effect the system operation?  This paper represents a first step in answering these questions using individual microcontrollers to emulate p-bits described by Eq.~(\ref{eq:psl3 p-bit}), while the interconnections described by Eq.~(\ref{eq:psl3 weight}) are implemented by another microcontroller. 

\par Our approach is quite similar to ref. 10, where electronic versions of synapses and neurons are built using off-the-shelf technology to demonstrate experimentally the formation of associative memory in a simple neural network consisting of three electronic neurons connected by two memristor-emulator synapses. Clearly our microcontroller based emulation of p-bit networks is not very scalable. But we envision that the interconnect between stochastic p-bits can be efficiently built using contemporary CMOS solutions and that nanodevices would be needed to build more efficient stochastic p-bits. This work primarily motivates such an endeavor and we develop essential rules of operation for such future systems. 

\par While the long term goal is to develop miniature integratable devices,
the hardware emulation presented here has many of its important features. The variables $\rm m_i(t)$ and $\rm I_i(t)$ appearing in Eqs.~(\ref{eq:psl3 p-bit}) and (\ref{eq:psl3 weight}) are not symbols represented in software, but actual voltages that can be observed and measured with oscilloscopes and voltmeters. The variability in the operation of real p-bits can be included by programming each microcontroller to have a different  retention time, $\tau_N$. Interconnect delays can be included into Eq.~(\ref{eq:psl3 weight}) as desired. The hardware implementation also allows us to establish important hardware rules for “interconnect delays” and retention times of p-bits, by systematically varying these time-constants.

\par Note that hardware implementations of Boltzmann Machines exist where Eq.(\ref{eq:psl3 weight}) is implemented in dedicated hardware while Eq.~(\ref{eq:psl3 p-bit}) has been simulated off chip. Both Eqs.~(\ref{eq:psl3 p-bit},\ref{eq:psl3 weight})have have been used as basis for dedicated VLSI based hardware implementations that perform various combinatorial optimization problems \cite{yoshimura2016fpga,okuyama2016computing} as well as hybrid architectures in context of learning \cite{kim2009highly,ly2009high,jarollahi2014nonvolatile,hu2015associative,ardakani2017vlsi,wang2017dlau} and combinatorial optimization \cite{bojnordi2016memristive}. This work, however, is focused on invertible Boolean logic, and is configured in a way that should be isomorphic with actual hardware implementations, where each microcontroller emulating a p-bit could be replaced with a specific hardware unit, such as a stochastic magnetic tunnel junction \cite{locatelli2014,majetich2016,grollier2016}, as we progress. 

\par To distinguish our PSL from other probabilistic logic concepts, it is necessary to put things into a historical context. The term “stochastic computing” or “probabilistic computing” has been used since 1960’s. The pioneering work of von Neumann \cite{von1956probabilistic}, Gaines \cite{gaines1969stochastic} and Poppelbaum et al. \cite{Poppelbaum:1967:SCE:1465611.1465696} addressed the reliable implementation of Boolean algebra and probabilistic arithmetic using stochastic components and established a field called “stochastic computing”. The major attraction of stochastic computing lies in its low complexity arithmetic units and inherent error tolerance.
\par A basic feature of stochastic computing is that numbers are represented by streams of bits that can be processed by simple circuits like AND gates, while the outputs are statistically counted as probabilities under both normal and faulty conditions. However, despite the advantages mentioned above, stochastic computing has been considered impractical because it takes a large number of bits to represent a value and does not show a cost advantage in multiplication – a prototypical inexpensive stochastic operation, when precision and reliability are required. Also the building block of such a system \cite{onizawa2016analog} will resemble some proposals \cite{camsari2016} of p-bits for PSL, but as we will describe in the next section, they are fundamentally different in their requirement to simultaneously read and write. An increase in the precision of a stochastic computation requires an exponential increase in bit-stream length, implying an exponentially increased computation time \cite{alaghi2013survey} \cite{manohar2015comparing}, which is undesirable. To be clear: We are not following this type of probabilistic approach but instead use a probabilistic architecture that offers substantial advantages over conventional computational schemes as described above.  

\par Next we describe the approach we are using to perform a hardware emulation of Eqs.~(\ref{eq:psl3 p-bit}) and (\ref{eq:psl3 weight}). Fig.~\ref{fig:p-bit} shows an emulation of a p-bit using a microcontroller. We then present a 3 p-bit Boltzmann Machine implementing an AND gate in both direct and inverted modes  of operation (Figs.~\ref{fig:AndGate1},\ref{fig:AndGate2}) and evaluate the role of sampling and retention times in ensuring proper operation (Figs.~\ref{fig:And_tau1},\ref{fig:Normalizedsampling},\ref{fig:And_tau2}). We then present results for binary adders in both direct and inverted modes (Figs.~\ref{fig:FullAdder},\ref{fig:4bitAdder}), and end with results for a 4-bit multiplier working in the inverted mode as a factorizer (Fig.~\ref{fig:factorizer}).

\section*{Methods}

\subsection*{Arduino pro mini as a p-bit}
A version of Eq.~(\ref{eq:psl3 p-bit}) suitable for microcontroller based emulation of a p-bit is given as

\begin{equation}
\rm V_{OUT}(t) = sgn\bigg\{ rand\big(-1,0\big) + S\big({V_{IN}}(t)\big)\bigg\} 
\label{eq:p-bit}
\end{equation}

\noindent where $\mathrm{V_{OUT}}$ and $\mathrm{V_{IN}}$ are the digital output and analog input voltages of the p-bit and S(x) is a sigmoidal function given by,
 
\begin{equation}
\rm S(x)=\frac{1}{1+e^{-2x}}
\label{eq:sigmoid}
\end{equation}

\noindent \textbf{I/O characteristics:} An Arduino pro mini is a 24 pin microcontroller \cite{arduinoRef} that can be programmed to emulate the behavior of Eq.~(\ref{eq:p-bit}) as shown in Algorithm 1. It has 6 dedicated analog input pins that have very high input resistances (100 M$\mathrm{\Omega}$) along with 6 dedicated PWM (Pulse-width modulation) output pins that have very low output resistances (100 $\mathrm{\Omega}$) with the ability to source 40 mA of current. This allows the Arduino to behave as a voltage controlled voltage source. 

\begin{figure*}[t]
\captionsetup[subfigure]{labelformat=empty}
\centering
\subfloat[]{\includegraphics[width=6.5in]{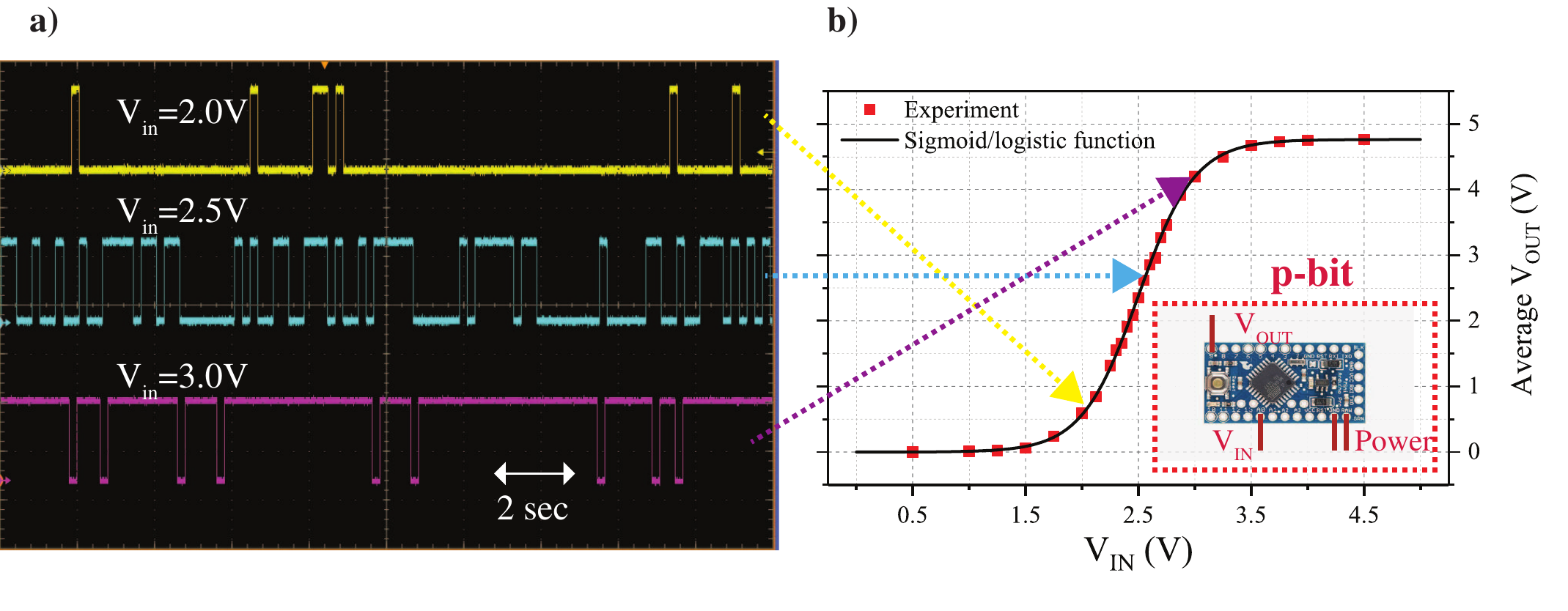}}

\caption{\textbf{p-bit emulated using Arduino microcontroller:}  Eq.~(\ref{eq:psl3 p-bit}) is emulated using the Arduino mini pro microcontroller as detailed in Algorithm 1. The microcontroller shown in the inset of (b) has dedicated analog input and digital output pins. The time evolution of the output voltages of p-bits is shown in (a) using a Tektronix DPO7104 oscilloscope. The p-bits produce more 1's than 0's as the input voltage is increased, demonstrating the tunable nature of the p-bit. Each of the red markers shown in (b) is a DC average measurement taken for a 100 second interval of the output voltage for a given input voltage. The average output voltages follows the sigmoidal function of Eq.~(\ref{eq:sigmoid}).}
\label{fig:p-bit}
\end{figure*}

\noindent \textbf{p-bit operation:} The time evolution of the output voltage for a set of input voltages using an oscilloscope (Tektronix DPO7104) is shown in Fig.~\ref{fig:p-bit}(a). As the input voltage is varied from low to high, the microcontroller generates more 1's than 0's. DC average measurements of the output voltage  taken  over 100 seconds are also shown  in Fig.~\ref{fig:p-bit}(b). The average voltage follows the sigmoidal function which indicates the tunable nature of the p-bit.

\par \textbf{Retention time $\mathrm{\tau_N}$:} Each p-bit is characterized by a retention time ($\mathrm{\tau_N}$) for which the output voltage is held constant. A possible physical component in the implementation of p-bits is the superparamagnet \cite{camsari2016}: 

\begin{equation}
\mathrm{\tau_N  = {\tau _0}\ {\exp \bigg( {\frac{\Delta}{kT}}\bigg)}}
\end{equation}

\noindent where $\mathrm{\tau_0}$ is a material dependent quantity ranging from  1 ps to 1 ns \cite{lopez2002transition}, $\Delta$ is the energy barrier of the nanomagnet and $\mathrm{kT}$ is the Boltzmann energy. For superparamagnets that are in the 10$-$20 $\mathrm{kT}$ range, the characteristic time is in the ms regime, assuming a $\tau_0$ of 1 ns. We emulate the retention time in our p-bits using a user defined delay $\mathrm{\tau_N}$ as shown in Algorithm 1. We later study the effect of retention time and establish some essential rules for proper operation of our interconnected p-bits.

\begin{algorithm}[H]
\caption{Pseudocode for p-bit}
  \label{alg:p-bit code}
    \begin{algorithmic}[H]
  
    \State \textbf{Parameters:} 
     \State \quad \quad \quad Digital output $\mathrm{V_{OUT}}$; 
     \State \quad \quad \quad Analog input $\mathrm{V_{IN}}$;
      \State \textbf{Repeat:}
      \State \quad \quad \quad  $x \gets$analogRead($\mathrm{V_{IN}}$);  \Comment{$\mathrm{V_{IN} \in (0,5 \ V), x \in (0.5)}$ }
      \State \quad \quad \quad  $m\gets 2x-5$;  \Comment{$\mathrm{m \in (-5,5)}$ }
      \State \quad \quad \quad  $\mathrm{Bias} \gets \mathrm{S(m)}$; \Comment{$\mathrm{Bias \in (0,1)}$ from Eq.(\ref{eq:sigmoid})      }
      \State \quad \quad \quad  $\mathrm{W} \gets \mathrm{rand(0,1)}$; \Comment{$\mathrm{W \sim U(0,1)}$ }
      \State \quad \quad \quad  \textbf{If}(Bias $>$ W)
      \State \quad \quad \quad \quad $\mathrm{V_{OUT}} \gets 1$; 
      \State \quad \quad \quad  \textbf{Else}						\Comment{$\mathrm{V_{OUT} \in \{0,5\ V\}}$ }
      \State \quad \quad \quad \quad $\mathrm{V_{OUT}} \gets 0$;
      \State \quad \quad \quad  \textbf{EndIf}
      \State \quad \quad \quad  \textbf{Wait} $\mathrm{\tau_N}$;
      \State \textbf{EndRepeat}
           
   \end{algorithmic}
\end{algorithm}

\subsection*{Weight Logic using microcontroller and DAC}

Fig.~\ref{fig:AndGate1}(a,b) shows a schematic and a block diagram for a 3 p-bit Boltzmann Machine that is programmed as an AND gate. The electrical wires connecting the components are not shown for clarity. The p-bits are correlated using a weight logic block that computes the input voltage of the $\mathrm{i^{th}}$ p-bit using the output voltages of all other p-bits in the network using

\begin{equation}
\mathrm{{V_{IN}}(t) = I_0\bigg({h_i} + \sum_j {{J_{ij}}{V_{OUT}}(t)}\bigg)} 
\label{eq:weight}
\end{equation}

\noindent where $\mathrm{\tau_{sample}}$ is the time interval for which the input voltages are held constant. Eq.~(\ref{eq:weight}) is a modified version of Eq.~(\ref{eq:psl3 weight}), meant to be used for our voltage controlled voltage source p-bits.

\par \textbf{Arduino mega as weight logic:} Our weight logic is implemented using an Arduino mega microcontroller in conjunction with MAXIM 5825 Digital to Analog converters \cite{maximDAC}. The Arduino mega can read as many as 52 digital inputs and communicates with the DAC using a fast $\mathrm{I_2C}$ protocol. The DAC has 8 channels with each having a 10-bit resolution. A pseudocode for programming an Arduino mega to emulate Eq.~(\ref{eq:weight}) is given in Algorithm 2. The input voltages of the p-bits set by Eq.~(\ref{eq:weight}) are not constrained in general, however we limit them to the p-bit input range between 0 and 5 Volts. Note that the weight logic not only correlates the p-bits, but can also be used for monitoring and recording the state of the Boltzmann Machines. Fig.~\ref{fig:AndGate1}(c,d,e) show two possible methods for monitoring the state of the system which are, 
\begin{itemize}

\item \textbf{Artificial nodes set through the DAC:} The microcontroller and the DAC can be used to create artificial voltage nodes that can be used to concurrently read the output  of the p-bits  as a single voltage. For example, in the operation of the AND gate ($\rm A \cap B = C$), 4$\times$A+2$\times$B+C  is evaluated and set as a voltage  in Fig.~\ref{fig:AndGate1}(c) to monitor the state of the AND gate.
\item \textbf{Serial logging:} The microcontroller that is part of the weight logic can also be used to log data through a serial port connection (USB). We have used this method extensively for collecting steady-state (long time) statistics for the various Boltzmann Machines that we present in this paper. 
\end{itemize}
\par Note that even though artificial nodes can be used to monitor the correlations of p-bits, serial logging of the data is much more convenient to collect long time statistics.  

\par \textbf{Communication between the DAC and Arduino mega:}  The  DACs use the $\mathrm{I_2C}$ protocol that allows the Arduino mega microcontroller to communicate with two pins SDA(Data) and SCL(Clock). When the system is first turned on, the DACs need to be initialized. This requires knowing the addresses of the individual DACs that are connected and setting a reference voltage for the DAC. We utilize at most 2 DACs within a Boltzmann Machine and the addresses for those are adjusted using two jumpers on the DAC. For example, to write a voltage of 2.5 V to channel 4 of the DAC whose address is set at ``0x20'', we could send the following 4 bytes over the $\mathrm{I_2C}$ interface: byte1 [0010000], byte2 [10110011], byte3 [10000000], byte4 [00000000]. The first byte has the address of the DAC in its 4 LSBs. The 4 MSBs of byte 2 has a command signal of writing to whichever channel is specified by the 4 LSBs of byte 2. The first 10 bits of byte 3 and 4 are the decimal equivalent of 512 which constitutes 2.5 V for a 10 bit DAC with 5V reference voltage. A library was written to internalize these operations, allowing the user to simply set voltages using a single write command that only uses the channel number and voltage for operation.
\begin{algorithm}[H]
\label{alg:weightlogic code}
  \caption{Pseudo code for  weight logic}
    \begin{algorithmic}[H]
  
      \State \textbf{Parameters:} 
      \State \quad \quad \quad Analog outputs $\mathrm{V_{IN}}$; \Comment{The input voltages of p-bits }
      \State \quad \quad \quad Digital inputs $\mathrm{V_{OUT}}$; \Comment{The output voltages of p-bits }
      \State \quad \quad \quad Parameters [J],$\rm \{h\}$ and $\mathrm{I_0}$; 
      \State \quad \quad \quad $n \gets$ Number of p-bits;
      \State \quad \quad \quad $k \gets$ DAC terminal for word;
      \State \textbf{Repeat:}

      \State \quad \quad \quad \textbf{For} i $\in\{1\cdots,n\}$ 
      \State \quad \quad \quad \quad $\mathrm{S} \gets$ digitalRead($\mathrm{V_{OUT}[i]}$); \Comment{ $\mathrm{V_{OUT} \in \{0,5V\}, S \in \{0,1\}}$ }
      \State \quad \quad \quad \quad $\mathrm{m \gets 2S-1}$ ; \Comment{$\mathrm{m \in \{-1,1\}}$ }
      \State \quad \quad \quad \textbf{EndFor} 
      
      \State \quad \quad \quad \textbf{For} j $\in\{1\cdots,n\}$ 
      \State \quad \quad \quad \quad Evaluate $\mathrm{{I_j}^{'} \gets I_0\big({h_j} + \sum_j {{J_{ij}}{m_j}}\big)}$ \Comment { $\mathrm{{I_j}^{'} \in (-\infty,+\infty)} $}
      \State \quad \quad \quad \quad \textbf{If}($\mathrm{{I_j}^{'}>5}$) 
      \State \quad \quad \quad \quad \quad $\mathrm{{I_j}=5}$ ;     		
      \State \quad \quad \quad \quad \textbf{ElseIf}($\mathrm{{I_j}^{'}<-5}$)
      \State \quad \quad \quad \quad \quad $\mathrm{{I_j}=-5}$ ; \Comment { $\mathrm{{I_j} \in (-5,+5)} $}
      \State \quad \quad \quad \quad \textbf{EndIf}
      \State \quad \quad \quad \quad $\mathrm{V_{IN}[j] \gets 2{I_j}-5}$
      \Comment { $\mathrm{{V_{IN}} \in (0,+5V)} $}
      \State \quad \quad \quad \quad Set DAC[j] $\gets \mathrm{V_{IN}[j]}$ 
      \State \quad \quad \quad \textbf{EndFor}
      
       \State \quad \quad \quad Set DAC[k] $\gets \mathrm{4\times V_{OUTA}+2\times V_{OUTB}+V_{OUTC}}$  \Comment{ Output word }
       \State \quad \quad \quad  Set Serial() $\gets \mathrm{V_{OUT}}$ for all p-bits \Comment { Output through the USB port}  
      
      \State \quad \quad \quad \textbf{Wait} $\mathrm{\tau_{D}}$ ;
          
      \State \textbf{EndRepeat}
           
   \end{algorithmic}
\end{algorithm}


\section*{Results}

\subsection*{AND Gate as a Boltzmann Machine}

\par
\begin{figure*}[t]
\centering
\captionsetup[subfigure]{labelformat=empty}
\subfloat[]{\includegraphics[width=6.5in]{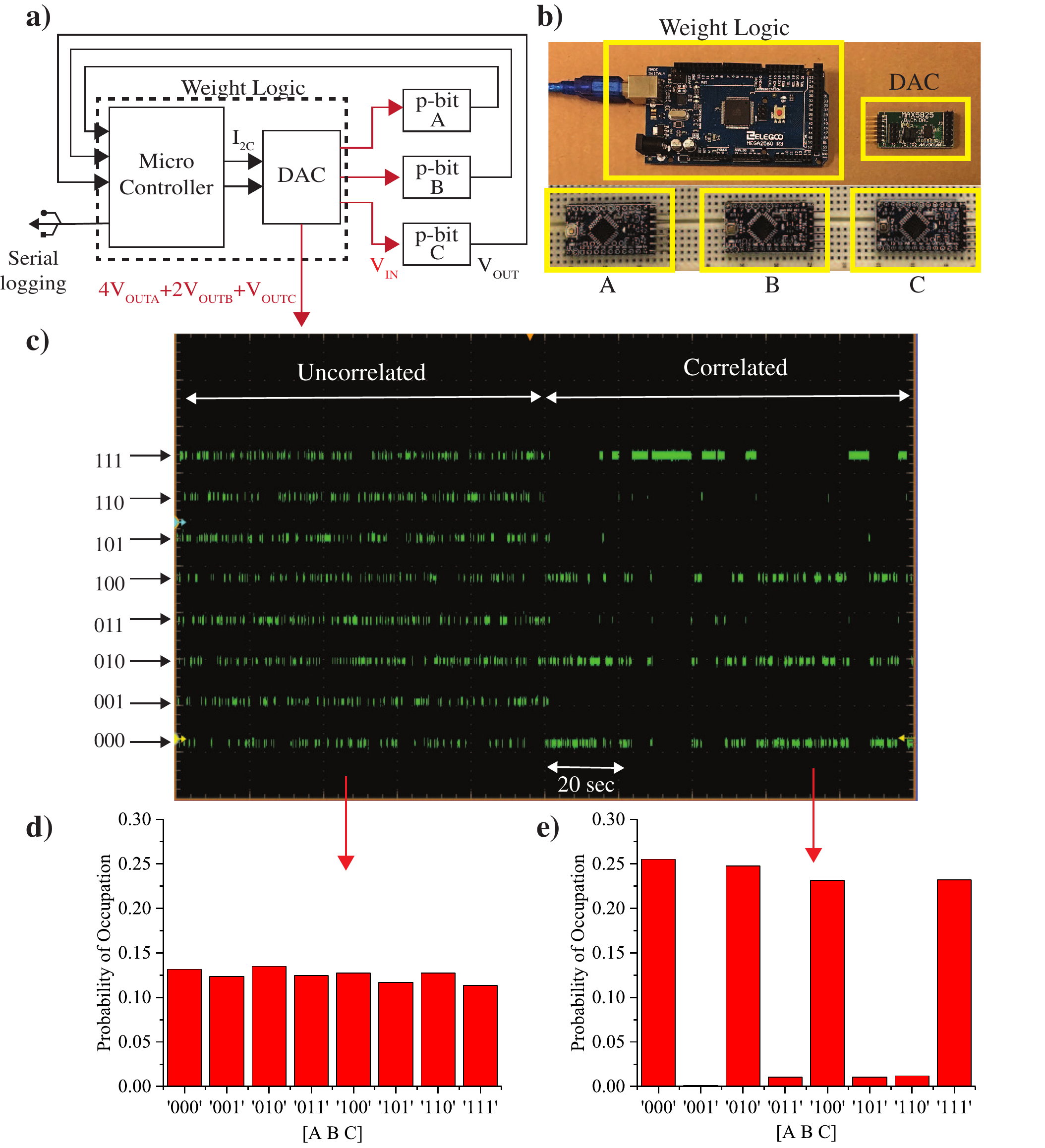}}

\caption{\textbf{AND gate constructed from 3 p-bits:} The block diagram and the schematic of an AND gate constructed using 3 p-bits are shown in (a) and (b). The electrical wires connecting the components are not shown for clarity. A weight logic block is used to correlate the p-bits as detailed in Algorithm 2. The output voltages of the 3 p-bits A, B and C are combined to form an artificial node 4$\times$A+2$\times$B+C, which is set using the DAC and is used to monitor the state of the system as shown in (c). As the system is left uncorrelated, it  goes through all possible 8 states of the artificial node with approximately equal probability. When the system is correlated using an $\mathrm{I_0=0.8}$, it visits the lines of the truth table with approximately equal probability. This is also seen by the steady-state statistics of the two cases presented in (d) and (e).}
\label{fig:AndGate1}
\end{figure*}

 \textbf{Correlated network of p-bits:} Fig.~\ref{fig:AndGate1}(c) shows the output voltage of an artificial node (4$\times$A+2$\times$B+C) as a function of time on the oscilloscope. For the AND gate the $\rm[J]$ and $\{h\}$  are taken from \cite{biamonte2008nonperturbative}. The strength of correlation between p-bits is adjusted through the parameter $\mathrm{I_0}$ in Eq.~(\ref{eq:psl3 weight}). $\mathrm{I_0}$ can be thought as the inverse (pseudo) temperature, in the sense that as $\mathrm{I_0}$ increases the p-bits get strongly correlated. When the system is  uncorrelated by using a $\mathrm{I_0}=0$, the 3 p-bits are independent of each other, resulting in the artificial node being uniformly distributed between 0 and 7, which can be seen from the steady state statistics for $\mathrm{I_0}=0$ as shown in Fig.~\ref{fig:AndGate1}(d). However, when the system is correlated using an $\mathrm{I_0}=0.8$, it locks to the states prescribed by $\rm [J]$ and $\rm \{h\}$ matrices, corresponding to the lines of the truth table for an AND gate which is shown by the steady-state statistics for $\mathrm{I_0}=0.8$ in Fig.~\ref{fig:AndGate1}(e). Note that we have left all the inputs and outputs floating, which results in all the lines of the truth table getting highlighted as $\rm I_0$ is increased. This ``floating'' mode of operation is a unique feature of correlated p-bits. The statistics shown in Fig.~\ref{fig:AndGate1}(d,e) have been collected through serial logging through the weight logic for up to half a million samples. 

\par \textbf{Clamping p-bits:} For Boolean computation,  the  p-bits need to be \textit{clamped} to produce a given output. This is done by simply connecting the input voltage of the p-bit to either ground or 5 V. This in essence  corresponds to applying a large bias, $\rm h_i$, to a given p-bit according to Eq.~(\ref{eq:weight}). A clamped p-bit operates on the corners of the sigmoidal response shown in Fig.~\ref{fig:p-bit}(b). Note that the input and output bits of a Boltzmann Machine are on an equal footing and can be clamped for direct and inverted operation respectively, as we discuss below.  

\par
\begin{figure*}[t]
\centering
\captionsetup[subfigure]{labelformat=empty}
\subfloat[]{\includegraphics[width=6.5in]{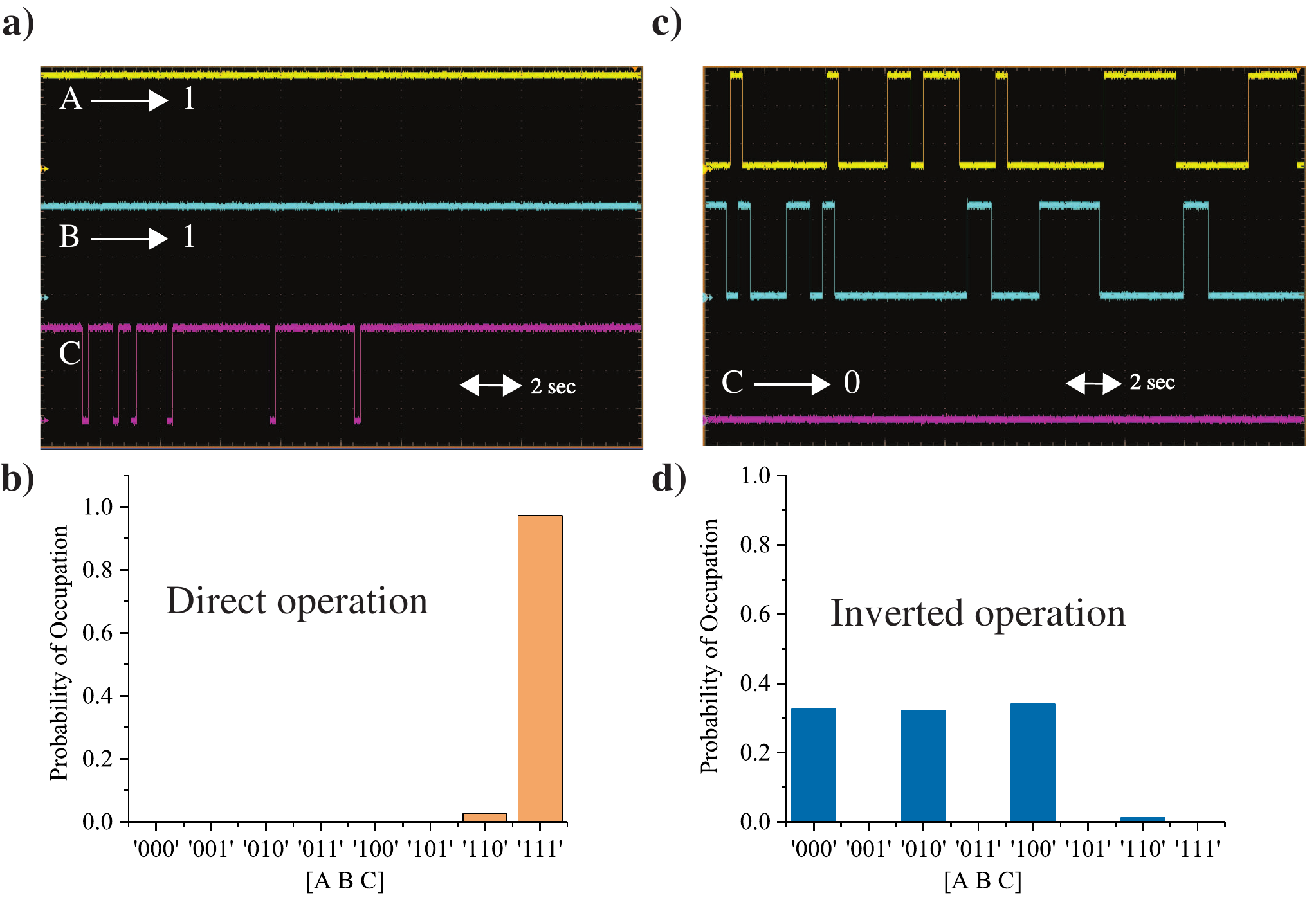}}

\caption{\textbf{Direct and Inverted operation:} (a) shows the time evolution of the output voltages of A, B and C on an oscilloscope. When A and B are clamped to 1, C mostly stays at  1 as shown in (a) and in the steady-state histogram shown in (b). (c) Remarkably, the system can operate in the inverted mode: When C is clamped to 0, the inputs A and B fluctuate between 00, 01 and 10, consistent for a C=0, with approximately equal probability as shown in the steady-state histogram in (d). }
\label{fig:AndGate2}
\end{figure*}

 \textbf{Direct Operation:} Fig.~\ref{fig:AndGate2} shows two cases of using an AND gate for computation purposes. Fig.~\ref{fig:AndGate2}(a) shows the time evolution of output voltages of p-bits A and B being clamped to 1 on the oscilloscope. As a result, the output voltages of C mostly stay in 1 as shown. This is also confirmed by the steady state statistics shown in Fig.~\ref{fig:AndGate2}(b) which are acquired using serial logging through the weight logic. 

\par \textbf{Inverted Operation:} A remarkable feature of the design is the \textit{inverted} operation. Fig.~\ref{fig:AndGate2}(c) shows the time evolution of output voltages for A, B and C when C is clamped to 0. It can be seen that A and B follow the states prescribed by lines of the truth table of an AND gate, as shown in Fig.~\ref{fig:AndGate2}(d). This feature stems from the fact that the system places all p-bits, whether input or output, on an equal footing. It is this inverted operation that can be used to solve more complex problems such as the 4-bit factorizer presented later in this paper.

\subsection*{Sampling and retention time}\label{sec:Samplingtime} 

Consider the Boltzmann Machine presented in Fig.~\ref{fig:AndGate1}. For each such network there are two major time constants: 

\begin{itemize}
\item Retention time $\mathrm{\tau_N}$ : Time interval for which the output voltage is held constant by the p-bit.

\item Sampling time $\mathrm{\tau_{sample}}$ : The time interval for which the input voltages to the p-bits are held constant by the weight logic. The sampling time can be thought of as the sum of the user defined delay $\mathrm{\tau_{D}}$ of Algorithm 2 and the time it takes to compute everything else in the Repeat block of Algorithm 2. 
\end{itemize}

\textbf{Boltzmann Law:} We now study the effect of both these time constants on the operation of the system using the AND gate. For such networks of correlated p-bits, an energy functional E for the state $\rm \{m\}=[m_i,m_j,\cdots] ^T $ can be defined as \cite{camsari2016}:

\begin{equation}
\mathrm{
E(\{m\}) =- I_0 \bigg(\sum_{i,j} \frac{1}{2} \left(J_{ij} m_i m_j\right) + \sum_i h_{i} m_i\bigg)}
\end{equation} 

The Boltzmann Law accurately captures the steady state probabilities of the system to be in different states $\rm\{m\}$ according to,

\begin{equation}
\mathrm{
P(\{m\}) = \frac{\exp(-E)}{ \sum_{i,j} \exp(-E)}
}
\label{eq:BL}
\end{equation}

\begin{figure*}[t]
\captionsetup[subfigure]{labelformat=empty}
\centering
\subfloat[]{\includegraphics[width=6.5in]{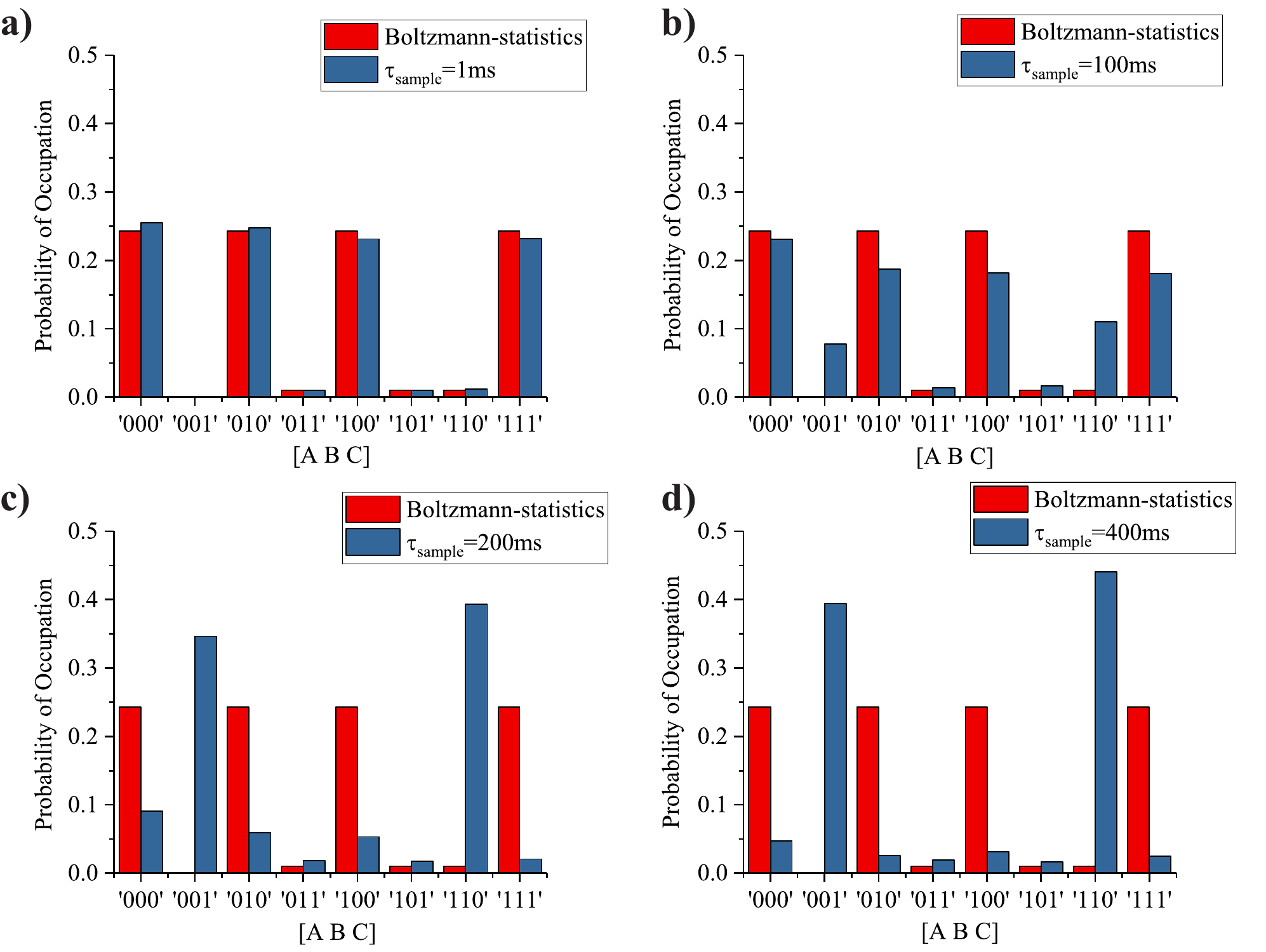}}

\caption{\textbf{Sampling time:} The sampling time $\mathrm{\tau_{sample}}$ is systematically varied from 1 ms to 400 ms while maintaining a  constant retention time of $\mathrm{\tau_N=200 \ ms}$ for each of the 3 p-bits. This is done by changing the user defined delay $\mathrm{\tau_D}$ in Algorithm 2. When $\mathrm{\tau_{sample}=1 \ ms}$, sampling is done much faster than the p-bit retention time and for this case steady-state statistics are well-described by the Boltzmann Law. As $\mathrm{\tau_{sample}}$ is increased to 100 ms  becoming comparable to the retention time of p-bits, two erroneous states 001 and 110 get highlighted more. When  $\rm \tau_{sample}$ is further increased to 200 ms the system completely breaks down. This trend repeates for $\rm \tau_{sample}$=400 ms.}
\label{fig:And_tau1}
\end{figure*}

 \textbf{Sampling time distribution:} Fig.~\ref{fig:And_tau1} shows the steady state statistics of an AND gate with each of the three p-bits having $\mathrm{\tau_N=200\  ms}$, with their sampling times $\mathrm{\tau_{sample}}$ varying from 1 ms to 400 ms. It can be seen from Fig.~\ref{fig:And_tau1}(a) that for extremely small $\mathrm{\tau_{sample}}$ the behavior of the system is captured well by the Boltzmann law. However as $\mathrm{\tau_{sample}}$ is increased to 100 ms, two incorrect states 001 and 110 stand out more. As $\mathrm{\tau_{sample}}$ is increased to 200 ms, the system  breaks down completely, with only the 001 and 110 states being highlighted. This continues for all $\mathrm{\tau_{sample}}$ greater then 200 ms as shown by $\mathrm{\tau_{sample}=400\ ms}$. 

\par
\begin{figure*}[p]
\captionsetup[subfigure]{labelformat=empty}
\centering
\subfloat[]{\includegraphics[width=\linewidth]{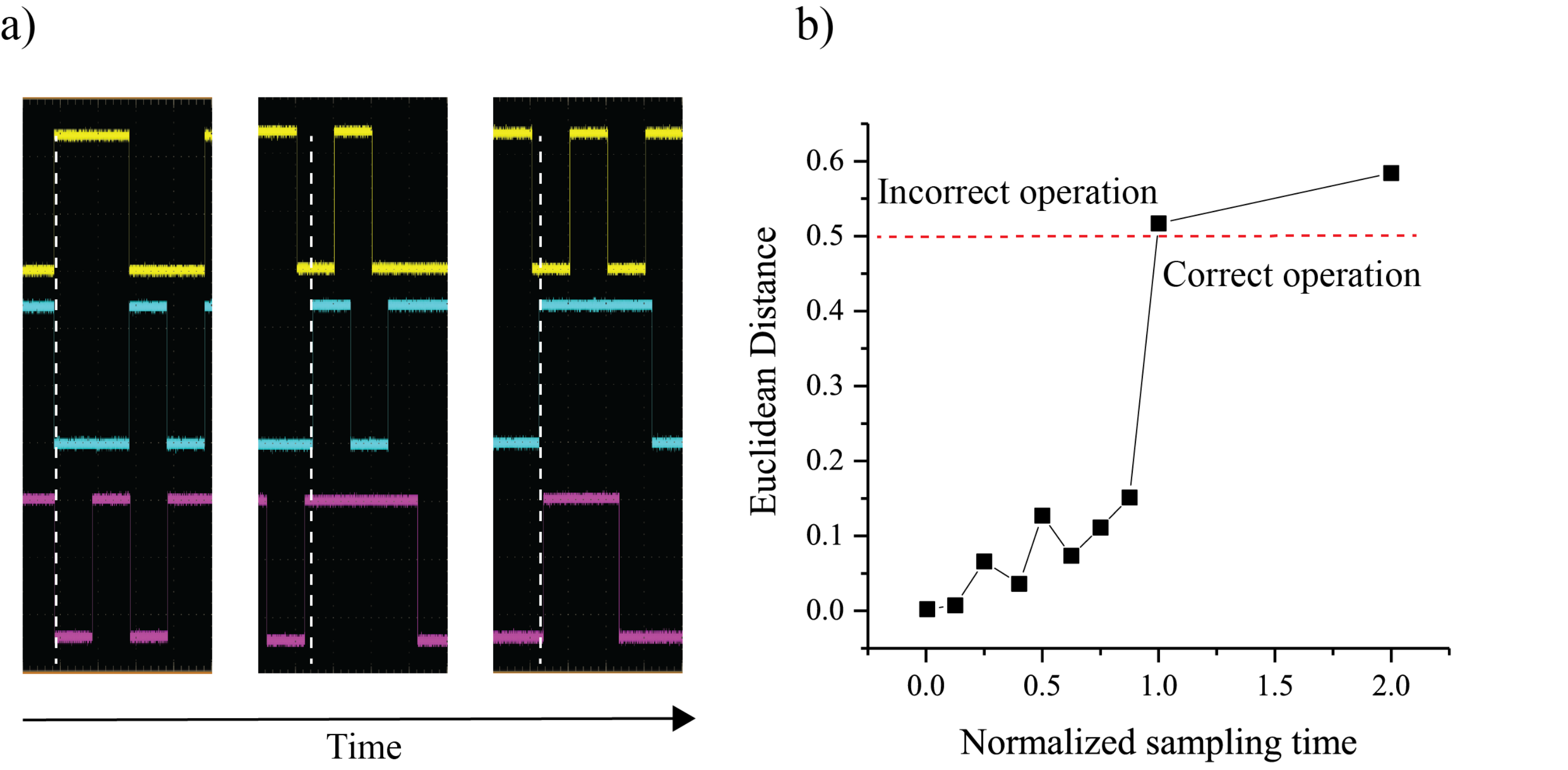}}

\caption{\textbf{\textit{Normalized} sampling time $\rm\big( \frac{\tau_{sampling}}{\tau_N}\big)$:} (a) An oscilloscope snapshot of the 3 p-bits is shown above as it changes as a function of time. Initially all 3 p-bits are almost perfectly aligned with extremely small phase differences between them. As time goes on this alignment is broken allowing each p-bit to update separately which leads to the system naturally having serial updates. (b) The euclidean distance between the steady state distribution of implemented hardware ( as a function of sampling time $ \rm \tau_{sampling} $ )  and the Boltzmann law is plotted as a function of \textit{normalized} sampling time. The sampling time $\rm \tau_{sampling} $ of the AND gate is varied from 1ms to 400ms, while the retention time of all p-bits is 200ms. For proper operation there is a hard threshold for the sampling time $ \rm \tau_{sampling} $ which is close to the retention time $ \rm \tau_N $ for a system with all p-bits having the same retention time. This condition reduces the probability of more then one p-bits getting updated simultaneously.  }
\label{fig:Normalizedsampling}
\end{figure*}

 We observe that when the sampling time is close to the retention time ($\rm \tau_{sample} \approx \tau_N$), Fig.\ref{fig:Normalizedsampling}(b) shows the euclidean distance between steady state distributions for various normalized sampling times ( sampling times from 1 ms to 400 ms with p-bit retention time of 200 ms). We observe that a boundary ($\rm \tau_{sample} \approx \tau_N$) exists for proper operation of the system. Around this boundary, p-bits can change their state before their input to the other p-bits are communicated, and this results in an incorrect operation. However, for fast sampling $\rm \tau_{sample} \gg \tau_N$, the updating is approximately instantaneous.  It is important to note that this requirement of $\rm \tau_{sample} \gg \tau_N$ necessitates a fast weight logic operation in any hardware implementation of p-bits.

\par An essential requirement for Hopfield networks and unrestricted Boltzmann Machines is the need for sequential updating, where each p-bit is updated serially but in any random order \cite{amit1992modeling,suzuki2013chaotic}, as opposed to parallel updating where each p-bit is updated at once. To enforce serial updating in asynchronous networks in simulation requires control flow statements which regulate the updating procedure of p-bits to one by one. Serial updating arises naturally in our setup since each p-bit is completely independent of each other and small phase differences that are present initially get greatly magnified as the system is run for longer periods of time, in the absence of a central clock signal. This type of updating is also known as the ``asynchronous dynamic'' in Hopfield networks \cite{amit1992modeling}. This is shown for an AND gate with 3 p-bits in Fig.\ref{fig:Normalizedsampling}(a), where each of the 3 p-bits are almost perfectly aligned to each other initially, however this alignment is broken as system continues to run with time. Asynchronous machines are known to converge slowly, while their synchronous counterparts allow for parallel updating, allowing much faster convergence. For hardware implementations, it is the synchronous Boltzmann Machines or Restricted Boltzmann Machines that would require some master control to ensure parallel updating making the system grow in resources as the number of p-bits increase.

\par
\begin{figure*}[p]
\captionsetup[subfigure]{labelformat=empty}
\centering
\subfloat[]{\includegraphics[width=\linewidth]{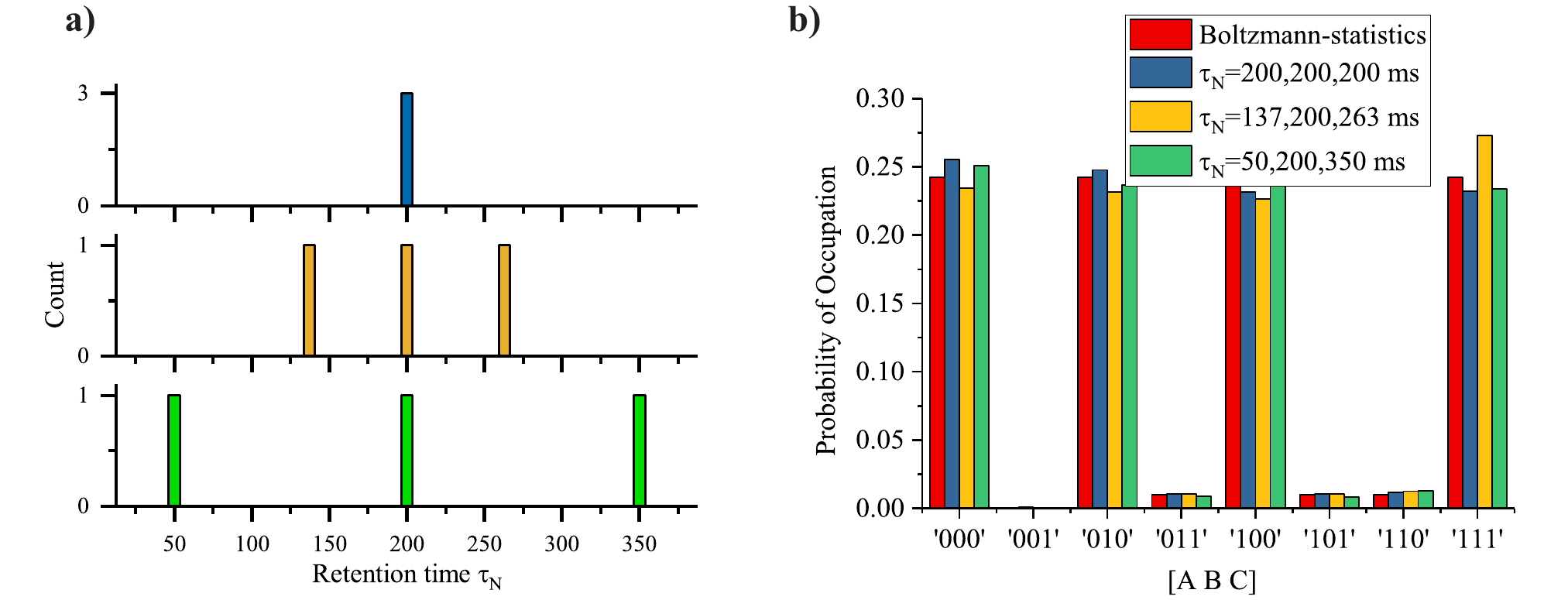}}

\caption{\textbf{p-bit retention time $\mathrm{\tau_N}$:} The retention time $\mathrm{\tau_N}$ of the p-bits is varied while maintaining a sampling time $\mathrm{\tau_{sample}=1 \ ms}$. This can be done by changing the variable $\mathrm{\tau_N}$ defined in Algorithm 1. In the most trivial case all 3 p-bits have the same $\mathrm{\tau_N=200 \ ms}$, while in the other two they are distributed over the mean as shown in histogram of $\mathrm{\tau_N}$ in (a) with a spread of $\pm33\%$ and $\pm75\%$. The steady-state statistics for each of the 3 cases shown in (b) are good matches with Boltzmann Law. This shows that as long as $\mathrm{\tau_{sample}}$ is much greater than $\mathrm{\tau_N}$ the system functions properly.   }
\label{fig:And_tau2}
\end{figure*}

 \textbf{Retention time distribution:} We now investigate the behavior of an AND gate in the presence of p-bits with different retention times that would arise due to inevitable process variations in a nanoscale  implementation. Fig.~\ref{fig:And_tau2}(a) shows the histogram for three different retention time configurations of the AND gate. In the most trivial case, all three p-bits have the same retention time $\mathrm{\tau_N=200 \ ms}$ while having a sampling time $\mathrm{\tau_{sample}=1\ ms}$. The steady state statistics for this case   exhibit a good match with the Boltzmann law (Fig.~\ref{fig:And_tau2}(b)). However, this configuration is unlikely in the case of any physical system where some distribution is to be expected due to process variations. 

\par A more realistic scenario is that of the 3 p-bits having different retention times. Fig.~\ref{fig:And_tau2}(a) shows two cases where p-bits are distributed in two sets of \{137, 200, 263\} ms  and \{50, 200, 350\} ms with a spread of  $\pm 33\%$ and $\pm 75\%$  around the mean value of 200 ms respectively,  while maintaining very fast sampling times of $\mathrm{\tau_{sample}=1 \ ms}$. Both cases show a good match with the Boltzmann Law (Fig.~\ref{fig:And_tau2}(b)). We  conclude that if the sampling time $\mathrm{\tau_{sample}}$ is much greater than the smallest $\mathrm{\tau_N}$, the system operation is well described by the Boltzmann Law, which can be attributed to the much reduced probability of parallel updating.

\subsection*{Full Adder as a Boltzmann Machine}\label{sec:Fulladder}

\begin{figure*}[t]
\captionsetup[subfigure]{labelformat=empty}
\centering
\subfloat[]{\includegraphics[width=6in]{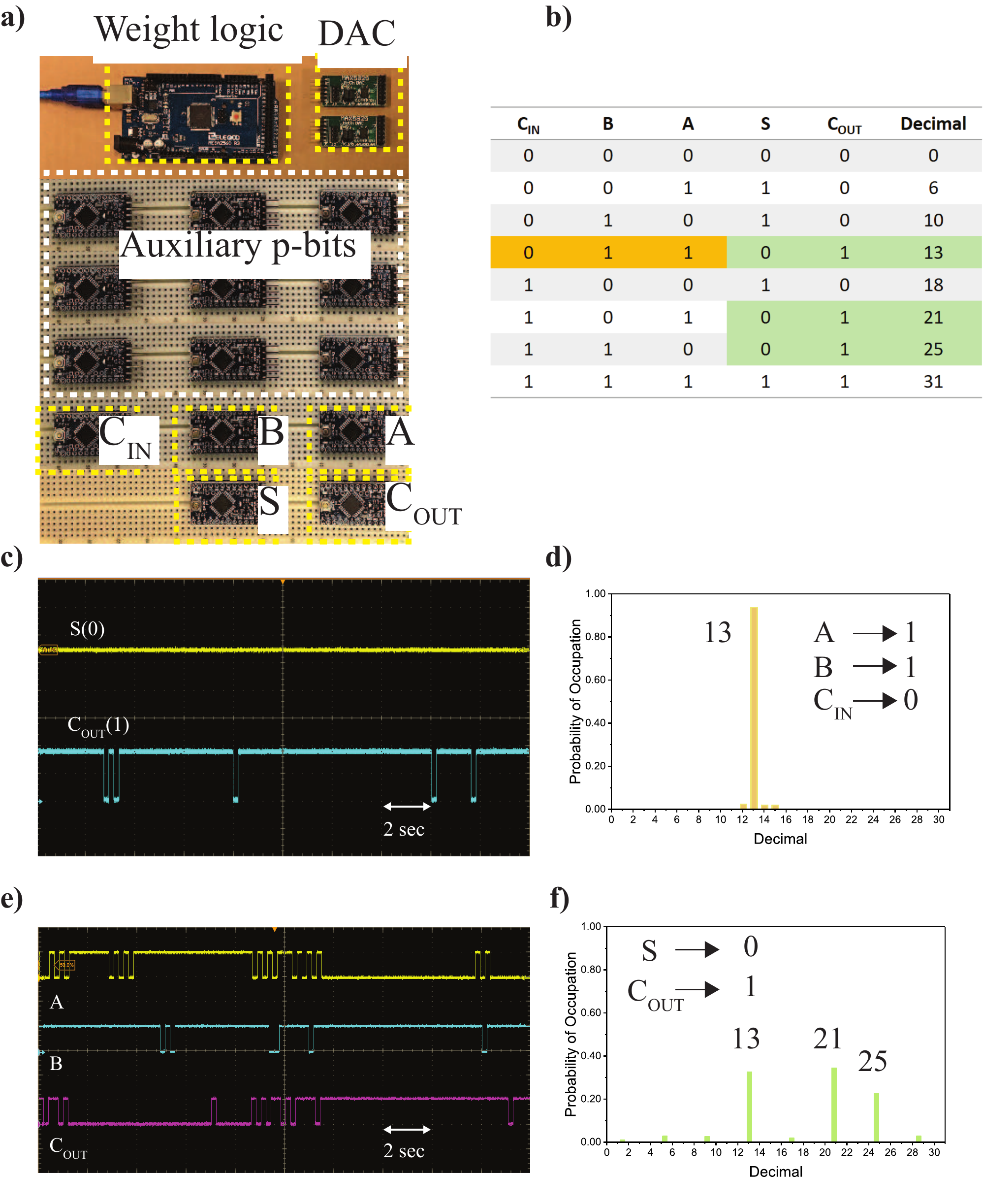}}

\caption{\textbf{Full Adder:} A Full Adder is implemented using 14 p-bits as shown in (a) along with its truth table in (b). The electrical wires connecting the components are not shown for clarity. (c) When the inputs $\mathrm{A}$, $\mathrm{B}$ and $\mathrm{C_{IN}}$ are clamped to 1,1 and 0 respectively, the Full Adder performs binary addition which can be seen from the time evolution of S and $\mathrm{C_{OUT}}$ on the  osilloscope. (d) The steady-state statistics acquired through serial logging are shown in (d). Since the Full Adder is bidirectional similar to the AND gate, the outputs $\mathrm{C_{OUT}}$ and $\rm S$ can be clamped to 1 and 0 respectively, that cause the inputs A, B and $\rm C_{IN}$ fluctuate among three states consistent with lines in the truth table as shown in (e) and (f).}
\label{fig:FullAdder}
\end{figure*}

 Fig.~\ref{fig:FullAdder}(a) shows a schematic of a 14 p-bit Full Adder implemented as a Boltzmann Machine. Of the 14 p-bits only 5 serve as the actual terminals of the Full Adder while the remaining 9 are  auxiliary p-bits. The retention and sampling times are chosen as $\mathrm{\tau_N=200\ ms}$ for all the p-bits with a $\mathrm{\tau_{sample}=10\ ms}$. However, now two DACs are needed to set the input voltages for all the p-bits since each DAC has 8-channels. 
 
\par The design of [J] and $\rm \{h\}$ matrices follows the treatment presented in \cite{camsari2016}. Direct computations can be performed by clamping p-bits as discussed earlier. Fig.~\ref{fig:FullAdder}(c,d) shows an example of 1-bit binary addition.The inputs A, B and $\mathrm{C_{IN}}$ have been clamped to 110 respectively, and the time evolution of output the voltages of S and $\mathrm{C_{OUT}}$ are shown in Fig.~\ref{fig:FullAdder}(c) which follow the states prescribed by the truth table of the Full Adder. This can also be seen from the steady state statistics shown in Fig.~\ref{fig:FullAdder}(d) which have been collected through serial logging.

\par Similar to the AND gate, the Full Adder implemented as a Boltzmann Machine can also be operated in inverted mode. The time evolution of the the inputs A, B and $\mathrm{C_{IN}}$ are shown in Fig.~\ref{fig:And_tau2}(e) when the outputs S and $\mathrm{C_{OUT}}$ are clamped to 0 and 1 respectively. The inputs A, B and $\mathrm{C_{IN}}$ follow the three prescribed states of the Full Adder truth table which is also confirmed by the steady state statistics shown in Fig.~\ref{fig:FullAdder}(f).

\subsection*{Directed Networks of Boltzmann Machines}\label{sec:Networks}

\begin{figure*}[t]
\captionsetup[subfigure]{labelformat=empty}
\centering
\subfloat[]{\includegraphics[width=6.5in]{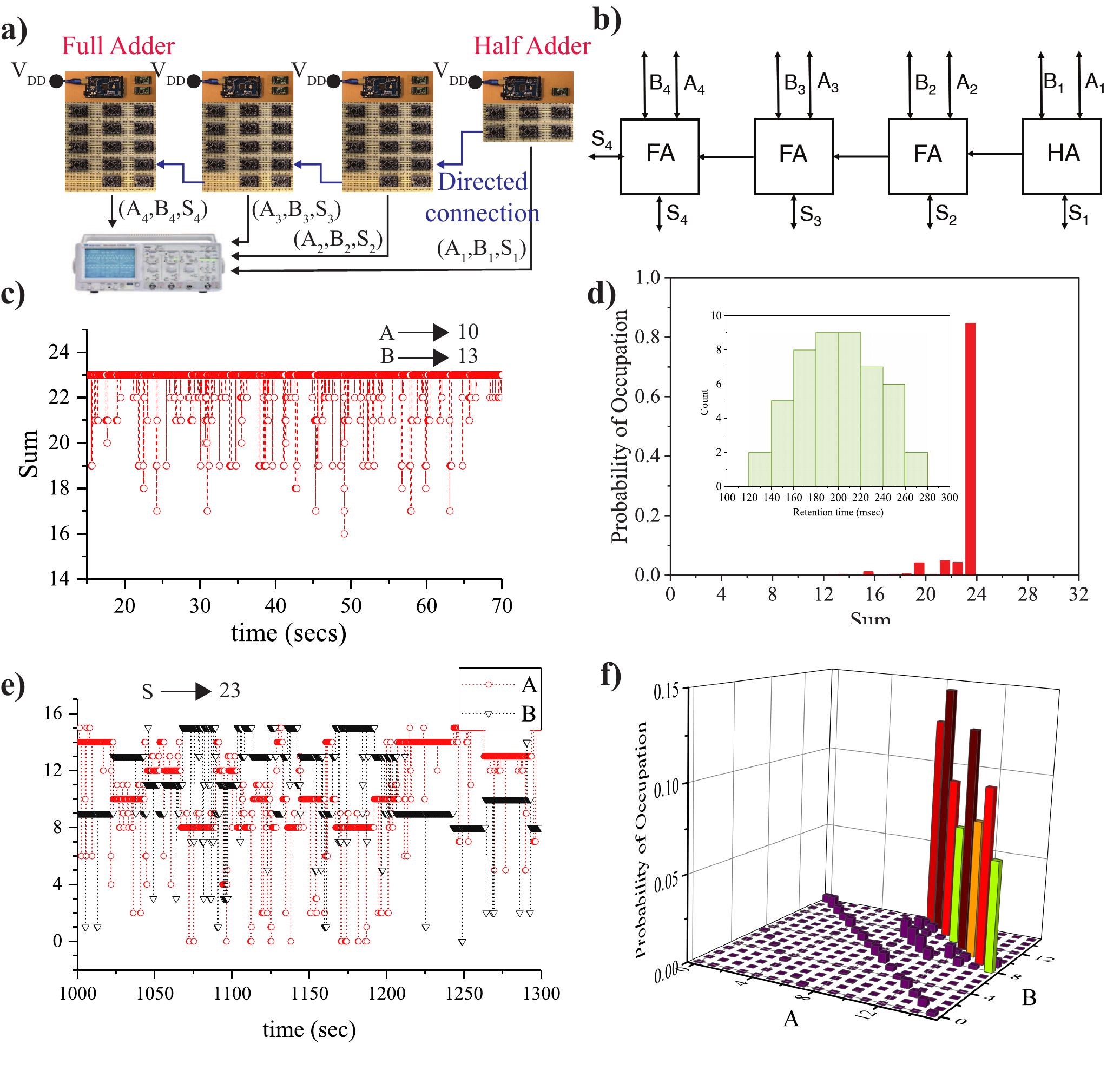}}

\caption{\textbf{4-bit Ripple Carry Adder (RCA)}: A 4-bit adder is implemented using 3 Full Adders and a Half Adder. A schematic and a block diagram are shown in (a) and (b). We assign each p-bit a separate retention time $\mathrm{\tau_N}$, with a normal distribution shown in the inset. (c-d) When the inputs are clamped to $\mathrm{A=10}$ to $\mathrm{B=13}$ the output S is 23. (e-f) In the inverted mode the output $\mathrm{S}$ is clamped to 23, resulting in $\mathrm{A}$ and $\mathrm{B}$ going through all 8 combinations (that can be probed by 4-digit binary inputs A and B) of producing A+B=S=23.}
\label{fig:4bitAdder}
\end{figure*}

\begin{figure*}[t]
\captionsetup[subfigure]{labelformat=empty}
\centering
\vspace{-5mm}
\subfloat[]{\includegraphics[width=0.9\linewidth]{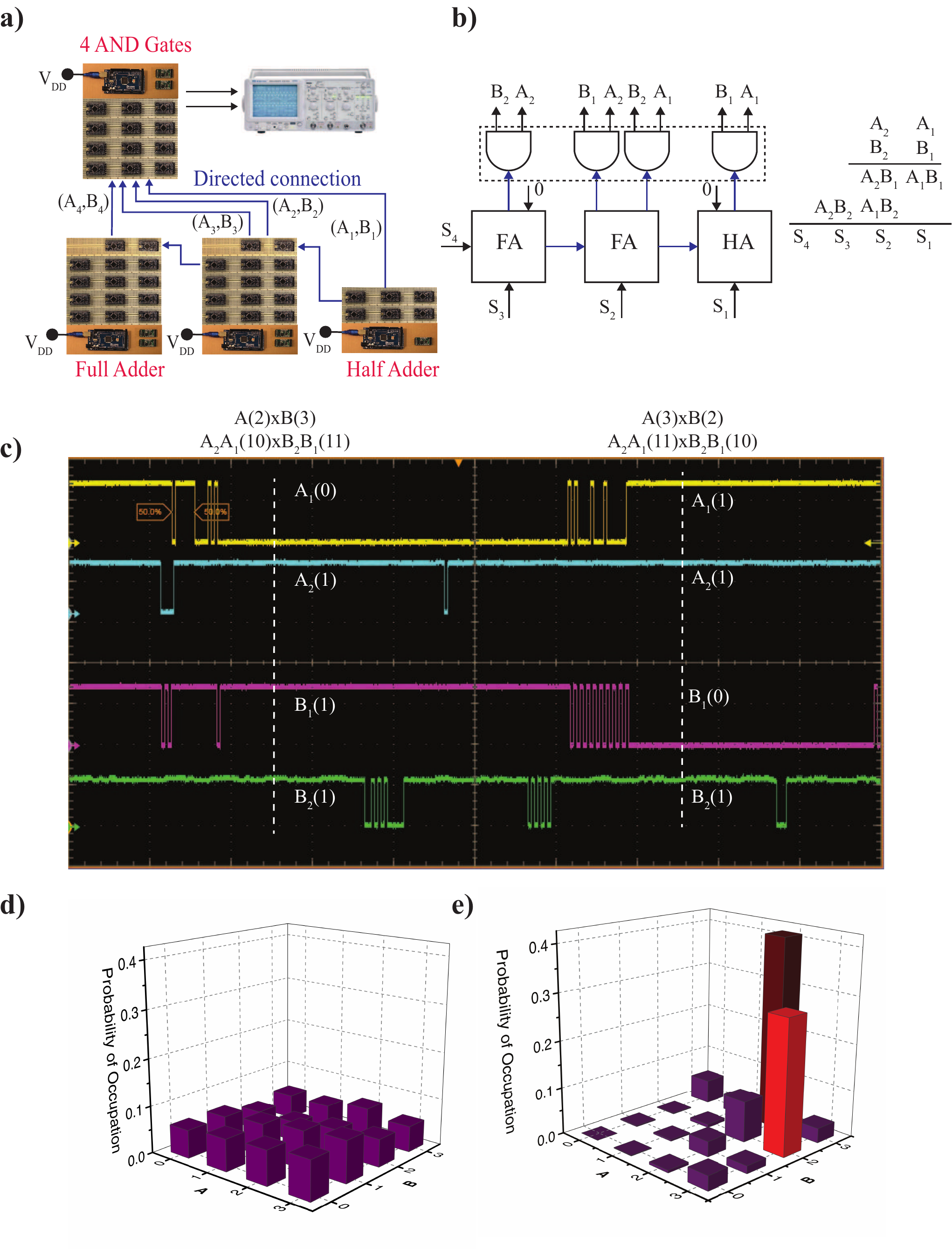}}

\caption{\textbf{4-bit multiplier/factorizer}: A 4-bit  multiplier constructed out of 3 Full Adders and 4 AND gates working in inverted mode operates as  factorizer. A  schematic and a block diagram are shown in (a) and (b). (c) When the sum of the 4-bit adder (product of the multiplier) is clamped to 6, the inputs A and B fluctuate between decimal 2 and 3 with approximately equal probability for the correlated system (e). For the uncorrelated system (f), the inputs fluctuate randomly among 16 possible states.}
\label{fig:factorizer}
\end{figure*}

To build more complex systems, one possible approach is to design the entire system as a single Boltzmann Machine, but the reversible nature of the Boltzmann Machines can hinder in the correct operation of such systems \cite{camsari2016}. A more practical alternative is to interconnect simpler Boltzmann Machines with \textit{directed} connections to build up more complex systems such as a 4-bit Ripple Carry Adder  (RCA) (Fig.\ref{fig:4bitAdder}(a)) or a 4-bit multiplier/factorizer (Fig.\ref{fig:factorizer}(a)). 

\par \textbf{Directed Connections:} Separate Boltzmann Machines can be connected in a \textit{directed} fashion such that the connections between the two are not reciprocal $\rm J_{ij} \neq J_{ji}$. In hardware, this corresponds to disconnecting the input voltage of  p-bit  ``i'' from its native weight logic and connecting to it the output voltage of  p-bit ``j'' from a different Boltzmann Machine so that $\rm J_{ij}=1 \  \mbox{and} \  J_{ji}=0$. Consider the case of a 4-bit adder that is built using a Half Adder and 3 Full Adders. In this case there are 3 directed connections as shown in Fig.~\ref{fig:4bitAdder}(a). Each connection takes the output voltage of $\mathrm{C_{OUT}}$ of the $\rm (n-1)^{\rm th}$ adder and connects it to the input terminal of $\mathrm{C_{IN}}$ of the  $\rm n^{\rm th}$  adder. Due to this connection scheme, no information can flow from the  $\rm n^{\rm th}$  adder to the  $\rm (n-1)^{\rm th}$  adder, which makes the system no longer bidirectional. However, as noted in \cite{camsari2016}, bidirectional connections of adders hinders proper operation of a n-bit adder. Also note that since the connection from one Boltzmann Machine to another is an electrical connection, the strength of the correlation between the two machines is at most 1 ($\rm J_{ji}=1$).  

\par \textbf{4-bit Adder:} We next demonstrate the correct operation of a 4-bit RCA comprised of 48 p-bits each having different $\mathrm{\tau_N}$ as shown in the inset of Fig.~\ref{fig:4bitAdder}(d). The values of $\mathrm{\tau_N}$ are normally distributed around an average of 200 ms with a minimum of 137 ms to a maximum of 263 ms, with a sampling time of 10ms for all Full Adders. 4-bit binary addition is performed by clamping the input p-bits of each adder, as demonstrated by the time evolution of the sum shown in Fig.~\ref{fig:4bitAdder}(c) with A=10 and B=13 resulting in the sum being 23 when converted to decimal. We observed for AND gates that their exists a boundary for proper operation of Boltzmann Machines with all p-bits having the same retention time. Similarly with a distribution such as the one studied here there also exists a boundary for proper operation which is $ \rm \tau_{sampling} \lessapprox min(\tau_N) $. This is due to the interconnect delays that need to be small.

\par \textbf{Inverted mode:} A more remarkable case is that of the sum bits of each of the adders being clamped to  S=23, with  A and B left floating. In this case, A and B fluctuate among  8 possible integer combinations  that satisfy A+B=23. Note that since A and B are 4-digit binary numbers, not all integer combinations can be probed by the system, for example A=22 and B=1.  This can be seen from the histogram presented in Fig.~\ref{fig:4bitAdder}(f).  Although there are 8 peaks in the histogram, the height of each peak is not the same since statistics presented in Fig.~\ref{fig:4bitAdder}(f) are not exactly steady state. With 48 p-bits in the system, the number of samples needed for steady state statistics is prohibitively large. Unrestricted Boltzmann Machines converge slower compared to restricted Boltzmann Machines\cite{hinton2007boltzmann}, but since asynchronous updates come naturally in hardware while synchronous updating will require more control circuitry, a design choice needs to be made between resources utilized and speed of convergence. Although it still remains to be seen how much of an improvement in the speed of convergence can be achieved by RBM's as compared to unrestricted Boltzmann machines.

\par  \textbf{4-bit multiplier/factorizer:} In this final example, we show how a standard digital multiplier built out of AND gates and Full Adders can be operated in reverse to function as a factorizer as shown in Fig.~\ref{fig:factorizer}, similar to what was proposed in \cite{traversa2017} in the  different context of memcomputing. Implementation of practically useful factorizers usually requires dedicated algorithms, here our purpose is simply to illustrate the remarkable invertibility of directed networks of p-bits. 
 
\par The block diagram of a digital multiplier is shown in Fig.~\ref{fig:factorizer}(b). The individual bits of A and B are first multiplied to produce $\mathrm{A_1B_1}$, $\mathrm{A_2B_1}$, $\mathrm{A_1B_2}$ and $\mathrm{A_2B_2}$ which are then added together to produce the product S. To convert this multiplier to a factorizer, we reverse the directed connections from the AND gates to the adders, while keeping the original directed connections of the Full Adders from the LSB to the MSB. 

\par The output voltages from the $\mathrm{A_X}$ and $\mathrm{B_X}$ (where X is the $\rm n^{th}$ Full Adder) are now sent as inputs to the output p-bits of the 4 AND gates. The 4 AND gates used here are part of one Boltzmann Machine instead of 4 separate Boltzmann Machines. This is because some inputs of the AND gates need to be the clones of each other as they go to different gates. For example, in Fig.~\ref{fig:factorizer}(b), $\mathrm{A_1}$ is a common input for the two right most AND gates, while $\mathrm{A_2}$ is a common input for the two left most AND gates. The retention and sampling times are chosen as $\mathrm{\tau_N=200\ ms}$ for all the p-bits with a $\mathrm{\tau_{sample}=100\ ms}$.  

\par Fig.~\ref{fig:factorizer}(c) shows the time evolution of output voltages of $\mathrm{A_1, B_1, A_2}$ and $\mathrm{B_2}$ using an oscilloscope when the sum of the adder is clamped to 6. This results in the input p-bits of the AND gates producing the correct factors of 3$\times$2 and 2$\times$3. This can also be seen by the statistics of the input p-bits of the AND gate as shown in Fig.~\ref{fig:factorizer}(e). As previously, the heights of both peaks are not the same due to the statistics not being exactly steady state. The results are collected through serial logging via the Boltzmann Machine for the AND gates. For comparison, we also show the statistics for an uncorrelated factorizer where 16 combinations are equally probable as shown in Fig.~\ref{fig:factorizer}(d).

\section*{Acknowledgments}
It is a pleasure to acknowledge many helpful discussions with Brian Sutton ( Purdue University). We are also grateful to Zhihong Chen (Purdue University) for discussions on stochastic computing. This work was supported in part by C-SPIN, one of six centers of STARnet, a Semiconductor Research Corporation program, sponsored by MARCO and DARPA, in part by the Nanoelectronics Research Initiative through the Institute for Nanoelectronics Discovery and Exploration (INDEX) Center, and in part by the National Science Foundation through the NCN NEEDS program, contract 1227020-EEC.

\section*{Author contributions statement}
Authors ( A.Z.P and L.A.G ) participated in conducting the experiments, while all authors ( A.Z.P, L.A.G, K.Y.C and S.D) helped in analyzing the results, reviewing and writing the manuscript.

\section*{Additional information}

\textbf{Competing financial interests} The authors declare no competing financial interests. 

\clearpage

\newpage

\end{document}